\documentclass[twocolumn,twoside,slac_two]{revtex4}
\usepackage{graphicx}
\usepackage{fancyhdr}
\begin{document}

\title{Parameters for a Super-Flavor-Factory}

\author{J.~T.~Seeman}
\author{Y.~Cai}
\author{S.~Ecklund}
\author{A.~Novokhatski}
\author{A.~Seryi}
\author{M.~Sullivan}
\author{U.~Wienands}
\affiliation{SLAC, 2575 Sand Hill Road, Menlo Park, CA 94025, USA}

\author{M.~Biagini} 
\author{P.~Raimondi}
\affiliation{INFN, Frascati, Italy}

\begin{abstract}
A Super Flavor Factory, an asymmetric energy $e^+e^-$ collider with a
luminosity of order 10$^{36}$ cm$^{-2}$s$^{-1}$, can provide a sensitive
probe of new physics in the flavor sector of the Standard Model. The
success of the PEP-II and KEKB asymmetric colliders 
\cite{KEKB,PEP_II} in producing
unprecedented luminosity above 10$^{34}$ cm$^{-2}$s$^{-1}$ has taught us
about the accelerator physics of asymmetric $e^+e^-$ colliders in a new
parameter regime. Furthermore, the success of the SLAC Linear Collider 
\cite{SLC}
and the subsequent work on the International Linear Collider \cite{ILC} allow a
new Super-Flavor collider to also incorporate linear collider
techniques. This note describes the parameters of an asymmetric
Flavor-Factory collider at a luminosity of order 10$^{36}$
cm$^{-2}$s$^{-1}$at the Y(4S) resonance and about 10$^{35}$
cm$^{-2}$s$^{-1}$ at the $\tau$ production threshold. Such a collider would
produce an integrated luminosity of about 10,000 fb$^{-1}$ (10 ab$^{-1}$)
in a running year (10$^{7}$ sec) at the Y(4S) resonance. In the following
note only the parameters relative to the Y(4S) resonance will be shown, the
ones relative to the lower energy operations are still under study.
\end{abstract}

\maketitle

\thispagestyle{fancy}

\section{Design from Past Successes}

The construction and operation of modern multi-bunch $e^+e^-$ colliders
have brought about many advances in accelerator physics in the area of high
currents, complex interaction regions, high beam-beam tune shifts, high
power RF systems, controlled beam instabilities, rapid injection rates, and
reliable uptimes ($\sim$95\%).

The present successful recent B-Factories have proven that their 
design concepts are valid:\\
1) Colliders with asymmetric energies can work. \\
2) Beam-beam energy transparency conditions are weak. \\
3) Interaction regions with two energies can work. \\
4) IR backgrounds can be handled successfully. \\
5) High current RF systems can be operated (3 A x1.8 A). \\
6) Beam-beam parameters can reach 0.06 to 0.09. \\
7) Injection rates are good and continuous injection is done in production. \\
8) The electron cloud effect (ECI) can be managed. \\
9) Bunch-by-bunch feedbacks at the 4 nsec spacing work well.

Lessons learned from SLC and subsequent linear collider studies 
(for ILC) and experiments (FFTB, ATF, ATF2) have also shown new 
successful concepts:\\
1) Small horizontal and vertical emittances can be produced in 
a damping ring with a short damping time. \\
2) Very small beam spot sizes and beta functions can be achieved 
at the interaction region. \\
3) Bunch length compression can be successfully performed.

All of the above techniques can be incorporated in the design 
of a future Super-Flavor Factory (Super-B) collider.

\section{Design Status}

The concept of combining linear and circular collider ideas to make a
linear-circular B-Factory was discussed in the late 1980's, although only
circular B-Factories were built in the 1990's. Recent advances in B-Factory
performance and solid linear collider design progress has reopened this
design avenue. The design presented here is very recent and on-going. The
parameters presented here are preliminary but with the intent to be
self-consistent.

\section{Luminosity}

The design of a 10$^{36}$ cm$^{-2}$s$^{-1}$ $e^+e^-$ collider combines
extensions of the design of the present \emph{B} Factories and linear
collider concepts to allow improved beam parameters to be achieved. The
luminosity L in an $e^+e^-$ collider is given by the expression
\begin{eqnarray*}
L      & = & \frac{N^+ N^- n_b f_c H_d }{4\pi \sigma_x \sigma_y } \\
\sigma & = & \sqrt{\beta \epsilon}
\end{eqnarray*}
where $n_b$ is the number of bunches, $f_c$ is the frequency of collision
of each bunch, $N$ is the number of particles in the positron (+) and
electron (-) bunches, $H_d$ is the disruption enhancement factor from the
collisions, $\sigma$ is the beam size in the horizontal ($x$) and vertical
($y$) directions, $\epsilon$ is the beam emittance and $\beta$ is the beta
function (cm) at the collision point for each plane.

\section{Collider Concepts Studied at the First Super-B Workshop}

\label{sec4}
\begin{figure}[htbp]
\begin{center}
\includegraphics[width=80mm]{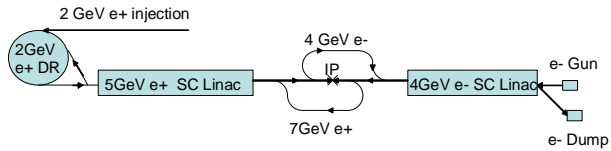}
\caption{Former Linearly Colliding Super-F Factory.}
\label{fig1}
\end{center}
\end{figure}

Schematic drawings of the Super-Flavor Factory as initially considered at
the First Super-B workshop \cite{LNF} is shown in Figure~\ref{fig1}. The operation is
described here. A positron bunch from a 2 GeV damping ring is extracted and
accelerated to 7 GeV in a superconducting (SC) linac. Simultaneously, an
electron bunch is generated in a gun and accelerated in a separate SC linac
to 4 GeV. The two bunches are prepared to collide in a transport line where
the bunch lengths are shortened. These bunches are focused to a small spot
at the collisions point and made to collide. The spent beams are returned
to their respective linacs with transport lines where they return their
energies to the SC accelerator. The 2 GeV positrons are returned to the
damping ring to restore the low emittances. The spent electron beam is
discarded. The process is repeated with the next bunch. It is expected that
each bunch will collide about 120 times each second and that there will be
about 10000 bunches. Thus, the collision rate is about 1.2 MHz. A small
electron linac and a positron source are used to replenish lost positrons
in the colliding process and natural beam lifetime. See Figure~\ref{fig1}.

\begin{figure}[htbp]
\begin{center}
\includegraphics[width=80mm]{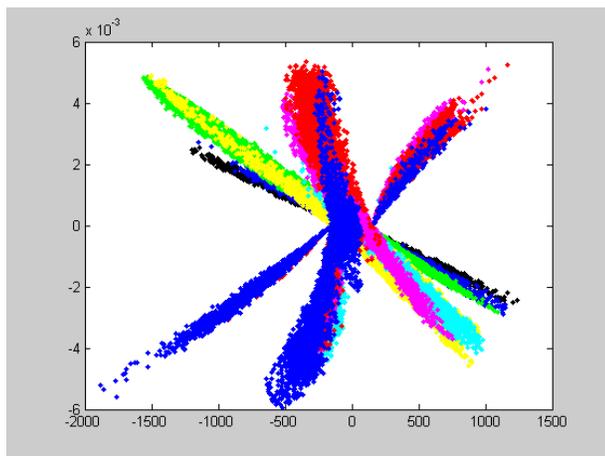}
\caption{Plot of the (y,y') phase space after collision in the
  earlier design. Each color refers to one longitudinal bunch slice.}
\label{fig2}
\end{center}
\end{figure}

This scheme was necessary in order to save power for cooling 
the beams that are heavily disrupted after the collision. As 
shown in Fig.~\ref{fig2}, the vertical emittance growth in a single collision 
is about 300. Running the rings at low energy is the only mean 
to bring the power requirements for the facility to the 100MW 
levels.

The scheme studied in \cite{LNF} presents several complexities and challenging
requirements for several subsystems. In particular the low energy required
for the rings, in combination with the high current, low emittance, small
energy spread and short bunch length, is more challenging than the already
challenging solution studied for ILC. Moreover, several technical solutions
proposed have never been tested and significant R\&D and detailed studies,
in order to ensure success, is required.

\section{Design Progress Presented at the Second Super-B Workshop}

The IP parameters have been re-optimized in order to minimize the
disruption due to the beam-beam forces. The proposed values, shown in 
Table~\ref{table1}, 
do produce a much smaller luminosity for a single pass, but the
emittance blowup for a single crossing is of the order of a few parts in
10$^{3}$ and, thus, only modest damping is needed between collisions. The
first column parameters are the best found so far for the previous
scheme. The large energy spread at the IP due to the bunch compression is
compensated with the monochromator scheme~\cite{Dubrovin}. Unfortunately, the
requirements are extremely ambitious; the bunch charge is much bigger than
the ILC one (a factor 3.5), the longitudinal emittance smaller (by factor
2) and the energy spread, together with the small $\beta_y$ at the IP is
incompatible with the Final Focus bandwidth.

\begin{figure}[htbp]
\begin{center}
\includegraphics[width=80mm]{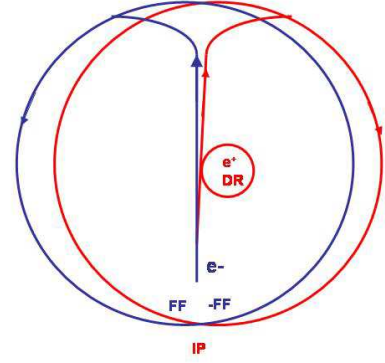}
\caption{Super-Flavor Factory layout.}
\label{fig3}
\end{center}
\end{figure}

\begin{figure}[htbp]
\begin{center}
\includegraphics[width=80mm]{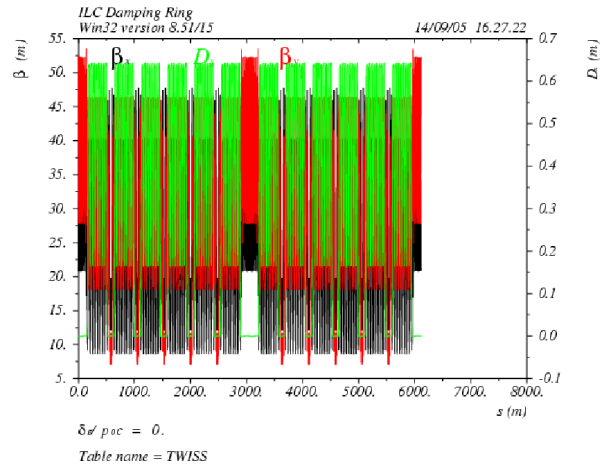}
\caption{ILC Damping Ring optical functions.}
\label{fig4}
\end{center}
\end{figure}

Fortunately, with the new scheme shown in Table~\ref{table1} 
column 2 and with such
small blowup, it is possible to increase the collision frequency, and
collide continuously in the ring with near ILC requirements. The proposed
parameters in the second column for the DR are nearly the ones proposed for
ILC except the number of bunches is about 4 times larger. The required
Final Focus is also exactly the one designed for ILC with the energy
rescaled. In Fig.~\ref{fig3} is shown the schematic layout. 
Fig.~\ref{fig4} shows the optical
functions of the ILC damping ring that operates at 5GeV and will be
rescaled to 4 and 7 GeV.

\begin{table}[h]
\begin{center}
\caption{Preliminary Super-F Factory collision parameters}
\begin{tabular}{|l|c|c|} \hline
 & 1$^\mathrm{st}$ LNF Workshop & 2$^\mathrm{nd}$ LNF Workshop \\
 & Best Working Point           & Best Working Point           \\ \hline
$\sigma_x^*$ ($\mu$m)      & 30 (1.0 $\sigma_x$
                                  betatron)     & 2.67         \\ \hline
$\eta_x$ (mm)              & 1.5 LER/--1.5 HER  & 0.0          \\ \hline
$\sigma_y^*$ (nm)          & 12.6               & 12.6         \\ \hline
$\beta_x^*$ (mm)           & 1.25               & 8.9          \\ \hline
$\beta_y^*$ (mm)           & 0.080              & 0.080        \\ \hline
$\sigma_z^*$ (mm)          & 0.100              & 6.0          \\ \hline
$\sigma_E^*$               & $2 \times 10^-2$   & $10^{-3}$    \\ \hline
$\sigma_{E\_\mathrm{Lum}}$ & $10^{-3}$          & $0.7\times10^{-3}$ \\ \hline
$\epsilon_x$ (nm)          & 0.8                & 0.8          \\ \hline
$\epsilon_y$ (nm)          & 0.002              & 0.002        \\ \hline
$\epsilon_z$ ($\mu$m)      & 2.0                & 4.0          \\ \hline
$\theta_x$ (mrad)          & Optional           & 2*20         \\ \hline
$\sigma_{z\_\mathrm{DR}}$
 (mm)                      & 4.0                & 6.0          \\ \hline
$\sigma_{E\_\mathrm{DR}}$  & $0.5\times10^{-3}$ & $10^{-3}$    \\ \hline
N$_\mathrm{part}$
 ($10^{10}$)               & 7.0                & 2.0          \\ \hline
N$_\mathrm{bunches}$       & 12000              & 12000        \\ \hline
I (A)                      & 6.7                & 1.9          \\ \hline
C$_\mathrm{DR}$ (km)       & 6.0                & 6.0          \\ \hline
$\tau_{x,y}$ (ms)          & 10                 & 20           \\ \hline
N$_\mathrm{turns}$
 bet. Coll.                & 50                 & 1            \\ \hline
f$_\mathrm{coll}$ (MHz)    & 12.0               & 650          \\ \hline
L$_\mathrm{singleturn}$
 ($10^{36}$)               & 1.5                & 1.2          \\ \hline
L$_\mathrm{multiturn}$
 ($10^{36}$)               & 1.1                & 1.0          \\ \hline
\end{tabular}
\label{table1}
\end{center}
\end{table}

\section{Collision Parameters Optimization}

Parameter optimization for this quasi-single pass factory has been
performed with the criteria described in Section~\ref{sec4}. 
Further studies and
ideas could relax the critical requirements, and they will continue. The
optimization for the ``collisions in the ring'' option is based on the
requirements to not have any need for bunch compression and
acceleration. The needs to have small IP spot sizes, small beta functions
and tune shifts are satisfied with the combination of small emittances and
a crossing angle. The low emittances reduce the beam sizes; the second one
simultaneously reduces the tune shift in both planes \cite{Shatilov} and the
longitudinal overlapping region. Since the interaction region now is short,
it is possible to decrease the vertical beta to very small values, further
decreasing the vertical size and tune shift. In addition, further
minimization of the beam-beam nonlinearities can be performed \cite{Raimondi}, 
to greatly reduce the residual emittance blowup due to the crossing angle.

\begin{figure}[htbp]
\begin{center}
\includegraphics[width=80mm]{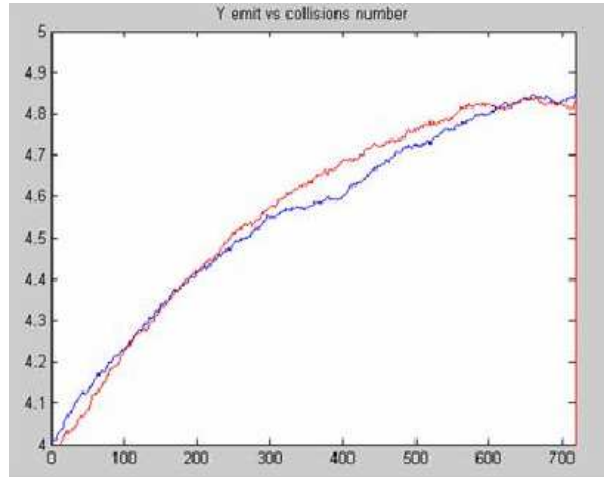}
\caption{Vertical emittance vs turn number.}
\label{fig5}
\end{center}
\end{figure}

Beam-beam studies have been performed with the ``GuineaPig'' computer code
by D. Schulte (CERN) \cite{Schulte}, 
which includes backgrounds calculations, pinch
effect, kink instability, quantum effects, energy loss, and luminosity
spectrum. This code has been intensely used for ILC studies of beam-beam
performances and backgrounds. In addition the code has been upgraded in
order to evaluate the equilibrium parameters when the collisions occur
repeatedly in a ring. The beams are tracked through the ring similarly to
what is done in \cite{Cai}, 
and the emittances and luminosity are evaluated after
equilibrium is reached. Fig.~\ref{fig5} shows an example of such tracking.

\section{Interaction Region Parameters}

The interaction region is being designed to leave about the same
longitudinal free space as that presently used by \emph{BABAR} but with
superconducting quadrupole doublets as close to the interaction region as
possible~\cite{Sullivan}.

Recent work at Brookhaven National Laboratory on precision conductor
placement of superconductors in large-bore low-field magnets has led to
quadrupoles in successful use in the interaction regions for the HERA
collider in Germany~\cite{Parker}. 
A minor redesign of these magnets will work well
for the Super F Factory.

A preliminary design of the Final Focus, similar to the NLC/ILC ones, has
been performed for the IP parameters in Table~\ref{table1}, 
second column. The total
FF length is about 70 m and the final doublet is at 0.5m from the IP. Such
a Final Focus needs to be inserted in one of the straight sections of the
ring.

\begin{figure}[htbp]
\begin{center}
\includegraphics[width=80mm]{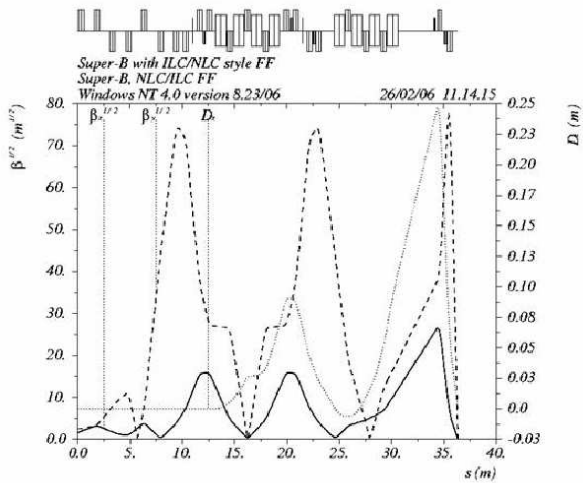}
\caption{Optical functions in half Final Focus region.}
\label{fig6}
\end{center}
\end{figure}

\begin{figure}[htbp]
\begin{center}
\includegraphics[width=80mm]{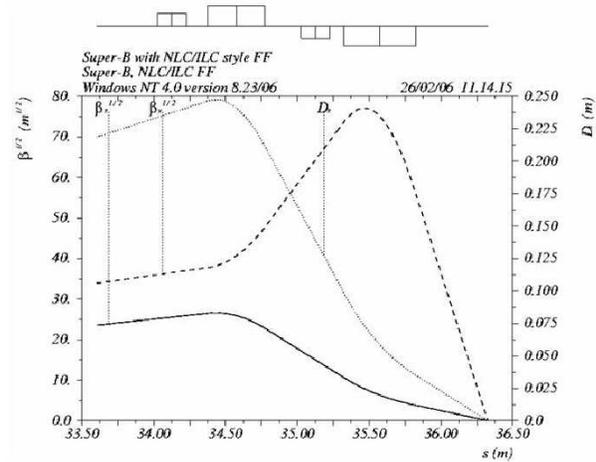}
\caption{Optical functions in the final doublet.}
\label{fig7}
\end{center}
\end{figure}

A plot of the optical functions in the incoming half of the FF region is
presented in Fig.~\ref{fig6}, optical functions in the final doublet is 
in Fig.~\ref{fig7}.

\begin{figure}[htbp]
\begin{center}
\includegraphics[width=80mm]{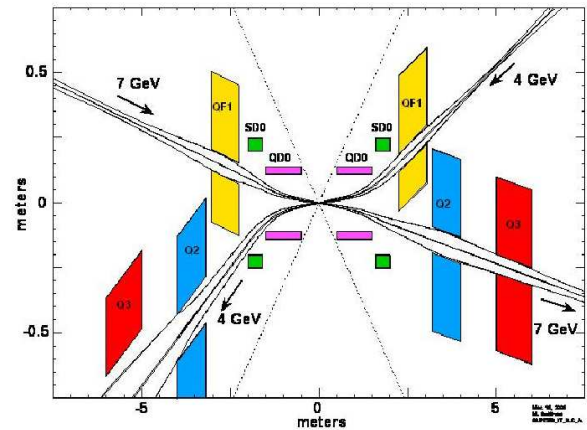}
\caption{Plan view of a possible IR layout.}
\label{fig8}
\end{center}
\end{figure}

The need for a finite crossing angle at the IP greatly simplifies the IR
design, since the two beams are now naturally separated at the parasitic
collisions. An expanded view of a preliminary IR layout is shown in 
Fig.~\ref{fig8}.

\section{Injector Concept and Parameters}

The injector for the Super Flavor Collider will make up for lost particles
with the finite beam lifetime in the damping rings and the losses from
collisions. The injector will be similar to the SLAC injector delivering
about 5 x 10$^{10}$ electrons or positrons per pulse at about 40 Hz each.

\section{Power Requirements}

The power required by a collider is the sum of a site base and the
accelerator operation. The damping ring power (about 40 to 60 MW) to
replace the synchrotron radiation loss will be the dominant factor in this
Super-B Factory. Better estimates and optimizations are under study.

\section{Synergy with ILC}

There are many similarities between this linear Super-B collider and the
ILC. The project described here will capitalize on R\&D projects that have
been concluded or are on-going within the ILC collaboration.

The damping rings between the two projects are now very similar. Main
differences are the ring energies (5 GeV for the ILC-DR, 4 GeV and 7GeV for
the Super-B) and the number of bunches (3000 for ILC, 12000 for
Super-B). For the ILC the bunch number is mainly dictated by the LINAC
requirements and the need to have enough spacing for the fast extraction
scheme. These requirements are not needed for the Super-F. Another critical
factor is the need to mitigate the electron cloud instabilities (ECI) in
the positron ring. This problem will be more severe in the Super-F. ECI
studies and R\&D have to be performed for both the ILC and Super-F to
mitigate this effect.

The interaction regions have very similar characteristics with flat beams
and overall geometries. The ratio of IP beta functions are similar (8-30 mm
horizontally and 0.08-0.5 mm vertically). The collimation schemes are
comparable. The chromatic corrections of the final doublets using
sextupoles will be the same. 

All beam bunches will need bunch-by-bunch
feedbacks to keep the beam instabilities and beam-beam collisions under
control. With the bunch spacing very similar, the feedback kickers, digital
controls, and beam impedance remediation will have common designs.

\section{Other Upgrade Possibilities}

The parameter optimization is continuously going on and we hope to further
reduce the criticality of several machine constraints. In addition more
careful studies are needed to make sure that the current constraints are
valid.

Additional improvements are also being considered for this design. In
particular we are studying the possibility to have a moderate bunch
compression in the ring that could relax some of the parameters like
emittances and beta functions.

The present scheme seems very promising but, given the rapid evolution of
the concepts, it might still have some weak points that can jeopardize
it. In addition new ideas and breakthroughs could also further change and
improve the design.

It has also to be pointed out that with the present scheme the Super-B
luminosity performance is a weak function with respect to the total length
of the ring. It has been chosen to be 6km mainly for the synergy with ILC,
but if there are strong constrains in terms of space and costs, it could be
reduced together with a re-optimization of the other parameters.

\bigskip % extra skip inserted
\begin{acknowledgments}
This document has come out of several recent Super-B workshops, with the
most recent one being at LNF (Frascati, Italy) on March 16-18, 2006. We
appreciate very much the useful discussions with the participants in these
workshops. We also appreciate the discussions of the parameters with
members of the ILC collaboration.

This work was supported in part by United States Department of Energy
contract DE-AC03-76SF00515.
\end{acknowledgments}

\bigskip % extra skip inserted

\end{document}